\documentclass[openacc]{rstransa}

\begin{document}

\title{Ice Giant System Exploration in the 2020s:  An Introduction}

\author{
L.N. Fletcher$^{1}$, A.A. Simon$^{2}$, M.D. Hofstadter$^{3}$, C.S. Arridge$^{4}$, I. Cohen$^{5}$, A. Masters$^{6}$, K. Mandt$^{5}$, and A. Coustenis$^{7}$}

\address{$^{1}$School of Physics and Astronomy, University of Leicester, University Road, Leicester, LE1 7RH, United Kingdom.\\
$^{2}$NASA Goddard Space Flight Center, Greenbelt, MD, 20771, USA.\\
$^{3}$Jet Propulsion Laboratory, California Institute of Technology, 4800 Oak Grove Drive, Pasadena, CA, 91109, USA.\\
$^{4}$ Department of Physics, Lancaster University, Bailrigg, Lancaster, LA1 4YB, UK.\\
$^{5}$Johns Hopkins University Applied Physics Laboratory, Laurel, MD, USA.\\
$^{6}$The Blackett Laboratory, Imperial College London, Prince Consort Road, London, SW7 2AZ, UK.\\
$^{7}$LESIA – Paris Observatory, CNRS, Paris Science Letters Research Univ., Univ. Paris-Diderot, Meudon, France.}

\subject{Planetary Science}


\corres{Leigh N. Fletcher\\
\email{leigh.fletcher@le.ac.uk}}

\begin{abstract}

The international planetary science community met in London in January 2020, united in the goal of realising the first dedicated robotic mission to the distant Ice Giants, Uranus and Neptune, as the only major class of Solar System planet yet to be comprehensively explored.  Ice-Giant-sized worlds appear to be a common outcome of the planet formation process, and pose unique and extreme tests of our understanding of exotic water-rich planetary interiors, dynamic and frigid atmospheres, complex magnetospheric configurations, geologically-rich icy satellites (both natural and captured), and delicate planetary rings.  This article introduces a special issue on Ice Giant System exploration at the start of the 2020s.  We review the scientific potential and existing mission design concepts for an ambitious international partnership for exploring Uranus and/or Neptune in the coming decades.

\end{abstract}


\begin{fmtext}






\end{fmtext}


\maketitle

\section{Introducing the Ice Giants}

It has been a mere 239 years, or $\sim8$ human generations, since Sir William Herschel's discovery of Uranus \cite{1781herschel}, and only 174 years ($\sim6$ generations) since theoretical predictions by Urbain Le Verrier \cite{1846leverrier} and John Couch Adams \cite{1846adams} were proven correct via Johann Gottfried Galle's \cite{1846galle} first observations of Neptune.  Since their discovery, Uranus has orbited the Sun 2.8 times, whereas Neptune has completed that journey only once.  For much of this time, these two distant worlds remained as wandering points of light against the fixed background of the cosmos.  Only with the coming of the space age, and specifically the spectacular Grand Tour of the Outer Solar System by Voyager 2 in the 1980s, did we begin to reveal the beauty and complexity of these rich planetary systems. With technology designed in the 1960s, and launched in the 1970s to formally explore the Gas Giants Jupiter and Saturn, Voyager 2 remains the only robotic explorer in history to fly past Uranus (24 January 1986) and Neptune (25 August 1989), offering only fleeting glimpses.  To this day, the Ice Giants Uranus and Neptune sit alone on the `Frozen Frontier', the only major class of planet yet to have a dedicated mission of exploration.  And yet their unique properties may hold the key to unlock the mysteries of planetary origins, both in our Solar System and beyond. 

This deep sense of perspective weighed heavily on the $200+$  participants who gathered at the Royal Society in London in January 2020, entering the Wellcome Trust lecture theatre past a display containing Herschel's hand-written notes on the discovery of Uranus.  This was the largest international gathering of Ice Giant scientists, engineers, mission planners, policymakers, and industry to date, with the purpose of revealing the scientific potential of new missions to explore the Ice Giants (their origins, interiors, atmospheres, and magnetospheres) and their rich planetary systems (their satellites, both natural and captured, and their rings).  Proposed for the 30th anniversary of Voyager's encounter with Neptune (1989), this Royal Society Discussion Meeting served to reinforce the growing momentum and international enthusiasm for an ambitious, paradigm-shifting mission to the Ice Giants as the next logical step in our exploration of the Solar System.  

The realm of the Ice Giants, between Uranus (19.2 AU) and Neptune (30.1 AU), remains largely unexplored.  Like the Gas Giants Jupiter and Saturn (which have been well-characterised by the Galileo, Juno, and Cassini orbital missions over the past three decades), these giant worlds accreted from the reservoirs of rocks, ices, and gases present in the protosolar nebula at the epoch of planet formation; they feature dynamic atmospheres with banded structures and localised storms;  convective interiors enriched in heavy elements (i.e., heavier than helium) compared to protosolar values; they exhibit powerful magnetic fields driven by hidden processes far below the clouds; and they are attended by delicate ring systems and ice-rich satellites.  But here the similarities end.  Uranus (14.5 Earth masses and 4.0 Earth radii) and Neptune (17.1 Earth masses and 3.8 Earth radii) represent an intermediate class of planetary object between the enormous Gas Giants and the smaller terrestrial worlds.  The expanding census of planets beyond our Solar System \cite{18fulton} has revealed that the most common outcomes of the planetary formation process are planets with radii intermediate between terrestrial worlds and the Ice Giants (primarily the `mini-Neptunes' between 1.8 and 4.0 Earth radii, that are not seen in our own Solar System - see Wakeford \& Dalba, this issue).  However, we note that this census from the Kepler mission remains biased to close-in planets with orbital periods shorter than 100 days, and that planetary population statistics is a field of extremely active research.  Furthermore, being similar in radius to an Ice Giant does not necessarily imply having similar planetary properties or environmental conditions.  Nevertheless, models of planetary origins, constrained by measurements of Ice Giant bulk composition (Uranus and Neptune seem to contain a much higher proportion of heavy elements compared to the hydrogen-rich giants) and internal structure, are challenged to explain how worlds of this size come to be, both in our Solar System, and beyond.  Uranus and Neptune therefore represent a ``missing link'' in our understanding of planet formation.

Uranus and Neptune exhibit stark differences, making them endmembers of this planetary classification, the products of divergent evolution from shared origins.  Many of their differences may be related to gargantuan collisions in their early history \cite{86stevenson}, with Uranus knocked entirely onto its side to experience the most extreme seasons in our Solar System (a $98^\circ$ obliquity, compared to Neptune's $28.3^\circ$).  Uranus' seemingly-sluggish atmosphere, with fewer storms and other meteorological phenomena than distant Neptune, may be a consequence of its negligible internal heat source, possibly related to the early collisions, or to the trapping of energy by stable layers within its interior.  The potentially water-rich interiors, which could be likened to global icy oceans, may help to explain why Ice Giant magnetospheres are unlike those found on any other Solar System object, with magnetic dipole axes wildly offset from the rotation axes, generating highly unusual time-variable interactions between the magnetosphere's internal plasma sources and the external solar wind.  Finally, the two worlds host completely different families of satellites and rings - Uranus features a natural, primordial satellite system, whereas Neptune is host to a captured Kuiper Belt Object (Triton), with active plumes and potential cryovolcanism, alongside the potential debris of an earlier, primordial satellite system (Nereid, Proteus, and the rings).  The Voyager flybys barely scratched the surface, and the two Ice Giant Systems therefore offer an ``embarrassment of riches'' for planetary scientists.

This special issue of \textit{Philosophical Transactions of the Royal Society A} results from the presentations, posters, and panel discussions at the \textit{Future Exploration of Ice Giant Systems} Discussion Meeting hosted by the Royal Society on January 20-21 2020.  This was immediately followed by parallel meetings hosted at the Royal Astronomical Society (\textit{Ice Giant Planets: Atmospheres, Origins and Interiors}) and Geological Society (\textit{Ice Giant Systems: Magnetospheres, Auroras, Rings and Satellites}) on January 22, 2020.  A summary of the meeting and the presentations is available online\footnote{\url{https://ice-giants.github.io}}, and the meeting was discussed extensively on social media\footnote{\url{https://twitter.com/hashtag/IceGiants2020 }}.  This introductory article seeks to offer context for the articles in this special issue, by looking back at the previous decade of Ice Giant mission concepts, and ahead to the opportunities in the 2020s.  These articles explore the new scientific insights into Ice Giant Systems being revealed today, but each paper calls for an ambitious new mission to dramatically advance our understanding of these enigmatic objects. Even if we were to launch at the end of the 2020s, an arrival at Uranus or Neptune in the 2040s will mean that more than half a century will have elapsed since humankinds' first and only encounters with these worlds.  The coming decade is therefore essential to prepare for a paradigm-shifting mission of discovery to the Ice Giants, which would contribute to shaping planetary science for a generation.

\section{Science Themes for Future Exploration}

\begin{figure}[!ht]
\centering\includegraphics[width=\textwidth]{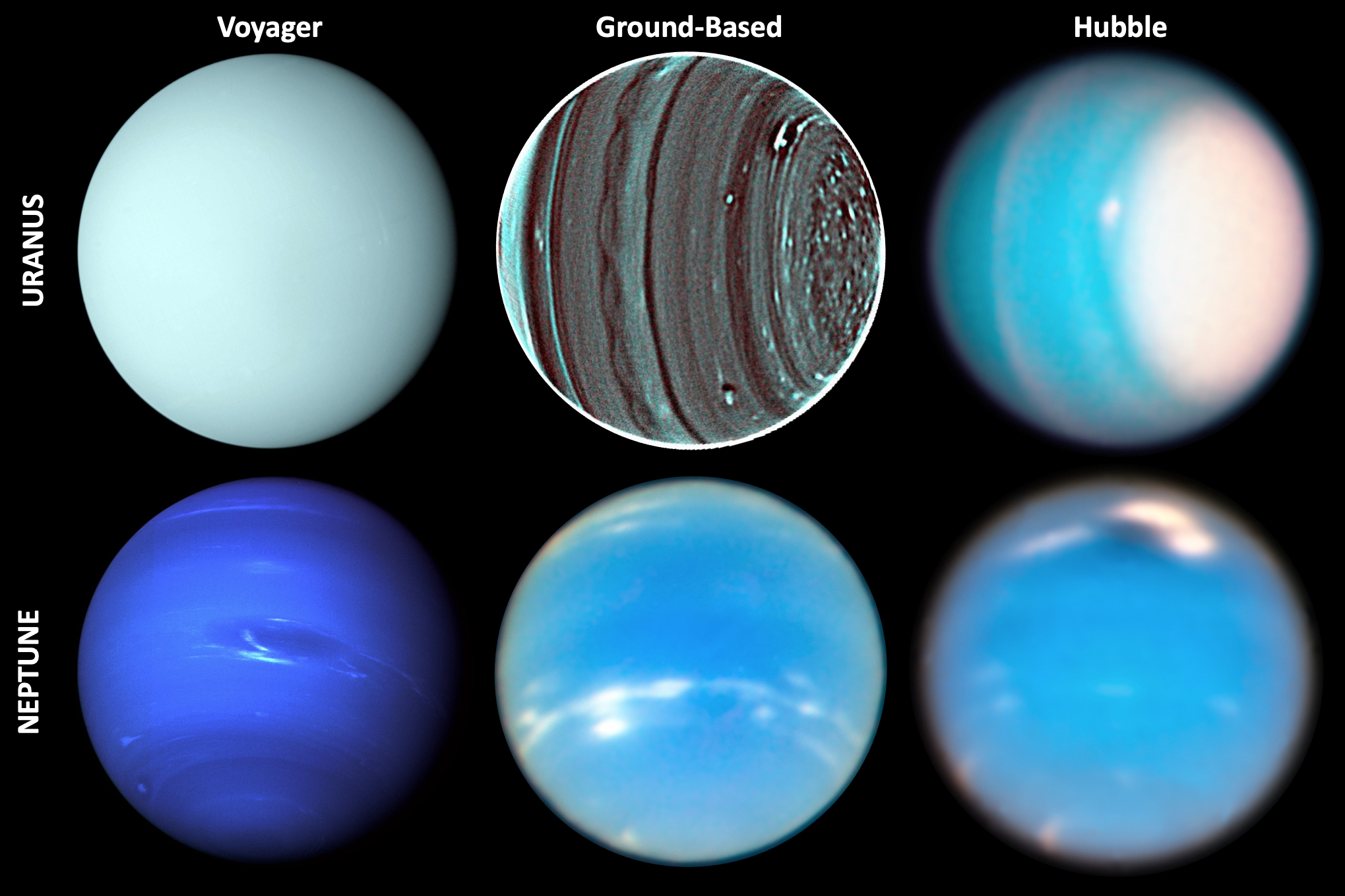}
\caption{The evolution of Ice Giant imaging from Voyager to the present day.  The left column represents our 20th-century views of Uranus (1986) and Neptune (1989) as observed by Voyager 2 (Credit:  NASA/JPL-Caltech).  The central column shows some of the best ground-based images of Uranus (near-IR from Keck in 2012, using high-contrast imaging to reveal the banded structure \cite{12fry}) and Neptune (visible-light from VLT/MUSE in 2018, Credit: ESO/P. Weilbacher (AIP).  The right column shows the Ice Giants observed by the Hubble Space Telescope in 2018 (Credit: NASA, ESA, A. Simon, M.H. Wong and A. Hsu).  Neptune's southern pole has been in view throughout this 30-year period, whereas Uranus was in southern summer in 1986 (left), and northern spring in 2018 (right). }
\label{montage}
\end{figure}

\begin{figure}[!ht]
\centering\includegraphics[width=\textwidth]{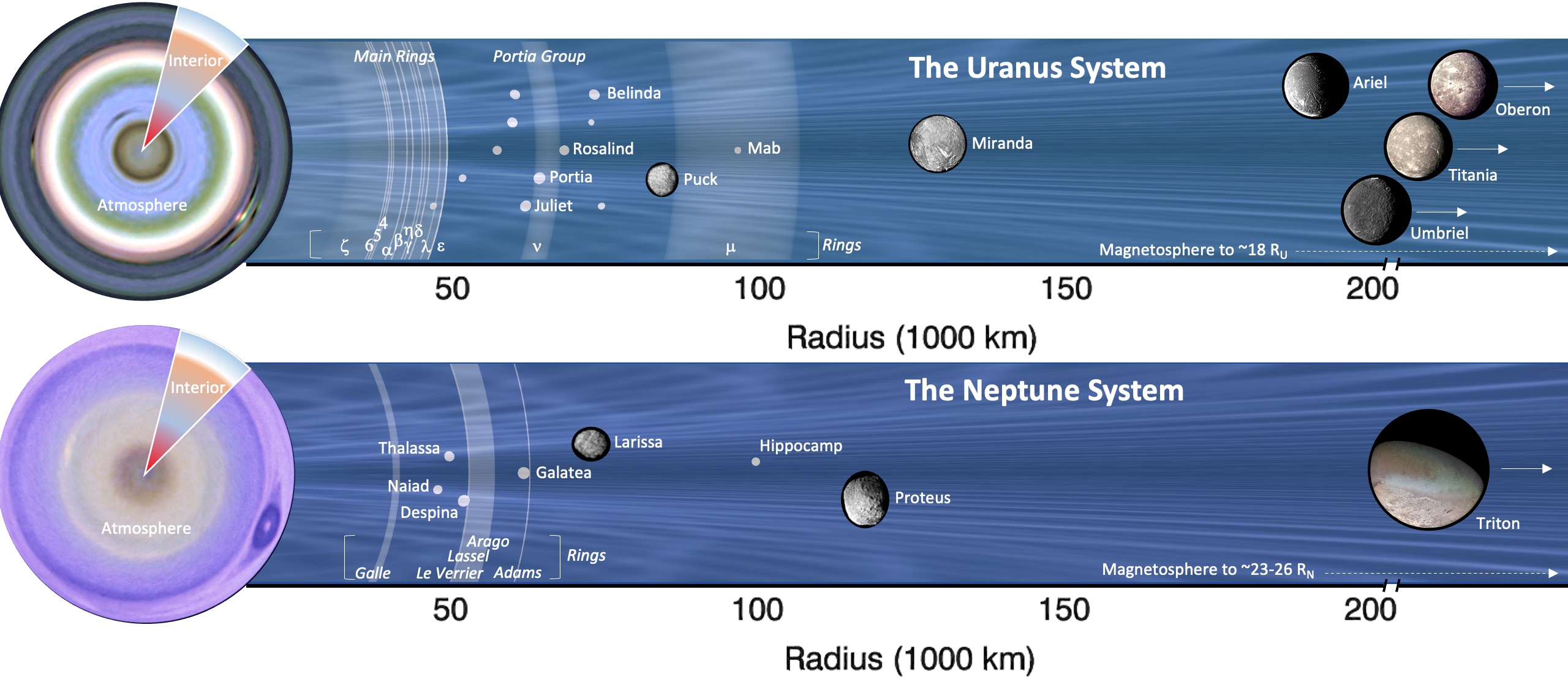}
\caption{The Ice Giant Systems offer a rich potential for discovery, from the deep interiors, dynamic atmospheres and complex magnetospheres of the planets themselves, to the delicate rings and myriad satellites that accompany each world.  The locations of rings and satellites in the inner systems are to scale with the planetary radii, with arrows next to major moons (right) indicating that they orbit at larger planetocentric distances.  Relative sizes of the satellites are not to scale.  Voyager images of Uranus and Neptune (left) are shown in false colour from a south-polar vantage point, with a cut-out representing the unknown density distribution of the interior.  The magnetosphere and radiation belts would encompass the full area of the figure.  Image modified from \cite{19fletcher_V2050}, Credit: L.N. Fletcher/M. Hedman/E. Karkoschka/Voyager-2.}
\label{system}
\end{figure}

Voyager 2 encountered both Ice Giants within a three-year ``golden era'' for planetary science between 1986 and 1989, summarised at the time by Stone \& Miner \cite{86stone, 89stone} in special issues of \textit{Science}, which also featured the first close-up images of both worlds by Smith \textit{et al.} shown in Fig. \ref{montage} \cite{86smith, 89smith}.  The wealth of data were scrutinised over the ensuing years, culminating in two substantial review books from the University of Arizona Press for Uranus in 1993  \cite{91bergstrahl}, and Neptune and Triton in 1996 \cite{96cruikshank}.  Voyager data remain the only source of \textit{in situ} measurements to date, and the only spatially-resolved observations of the Uranian and Neptunian satellite systems.  Nevertheless, the ensuing three decades saw ever-improving capabilities from ground- and space-based astronomical facilities, revealing new insights into the planets' atmospheres and ring systems.  Observations from an Earth-based vantage point are limited to the sun-facing hemispheres.  For example, observations have revealed the slow seasonal variability of Uranus' atmosphere (Fig. \ref{montage}) as the planet moved from southern summer solstice (1985), through northern spring equinox (2007), and will likely continue to change through northern summer solstice in 2030.  The long Neptunian year has meant that ground-based observations have been limited to Neptune's southern summer (solstice in 2005, with northern spring equinox in 2046).  These new observations, combined with advances in numerical simulations and the insights from Voyager, dominated the presentations at the Discussion Meeting (outlined below) and spanned the breadth of scientific potential revealed in Fig. \ref{system}.

\subsection{Planetary Atmospheres}

Spatially-resolved remote sensing in reflected sunlight (i.e., light scattered from atmospheric aerosols or rings) have relied on the Hubble Space Telescope in the UV, optical and near-IR \cite{09karkoschka, 11karkoschka_ch4, 14sromovsky}, and annual observations of both Uranus and Neptune are now part of the `Outer Planet Atmospheres Legacy' programme\footnote{\url{https://archive.stsci.edu/prepds/opal/}}, which provided the images in Fig. \ref{montage} (right column).  These have been greatly enhanced by the advent of ground-based telescopes with primary mirror diameters exceeding 8 m and good sensitivity in the near-IR (e.g., Gemini, Keck, VLT, etc.), capable of resolving the banded aerosol structure, zonal winds from tracking discrete cloud features, delicate gaps in the rings, and even the weak emission from the H$_{3}^{+}$ ion in Uranus' ionosphere.  Achieving adequate spatial resolution is more challenging at longer wavelengths, which provide details on atmospheric temperatures and gaseous composition:  observations from the Infrared Space Observatory \cite{02encrenaz}, Spitzer Space Telescope \cite{14orton}, Herschel Space Telescope \cite{10lellouch}, and AKARI Space Telescope \cite{10fletcher_akari} were all disc-integrated without spatial resolution.  More recent ground-based observations in the thermal-infrared \cite{07orton_nep, 14fletcher_nep, 15orton, 19roman} and millimetre (ALMA) and centimetre (VLA) range \cite{18depater, 19tollefson} have revealed the banded thermal structure of both worlds.  

Our knowledge of Ice Giant atmospheres was reviewed shortly after the Voyager encounters by Lunine \cite{93lunine}, and constraints on planetary origins from atmospheric composition and structure were reviewed by Mousis \textit{et al.} \cite{18mousis}.  These reviews have been recently updated: Hueso and Sanchez-Lavega \cite{19hueso} review atmospheric dynamics and winds; Fletcher \textit{et al.} \cite{20fletcher_icegiant} review Ice Giant banding and global circulation; and Aplin \textit{et al.} \cite{20aplin_icegiant} review measurements of atmospheric electricity on Uranus and Neptune.  Ice Giant atmospheres featured prominently on the first day of the Discussion meeting, and new insights are revealed in this special issue:  Hueso \textit{et al.} explore convective motions in hydrogen-rich atmospheres; Moses \textit{et al.} review the chemistry at work on the Ice Giants; and Melin \textit{et al.} and Moore \textit{et al.} describe the properties of the upper atmospheres and ionospheres.  Future exploration of atmospheric dynamics, circulation, chemistry, meteorology, and clouds would benefit from both orbital remote sensing and direct \textit{in situ} sampling, filling an unexplored gap in our understanding of planetary atmospheres in the weak-sunlight, low-temperature, and extreme-seasons regime.

\subsection{Origins and Interiors}

The properties of the hidden interiors of Uranus and Neptune may hold the secrets to the origin and evolution of these intermediate-sized worlds throughout our universe, as one of the most common outcomes of the planet formation process.  Mousis \textit{et al.} \cite{18mousis} and Atreya \textit{et al.} \cite{20atreya_icegiant} recently reviewed the deep atmospheric composition as a constraint on planet formation, and Helled \textit{et al.} \cite{20helled_icegiant} explored our knowledge of their internal structures.  Open questions on the structure and evolution of these two worlds are discussed in this special issue by Helled and Fortney, who focus on why the two planets may be different, rather than the same.  Teanby \textit{et al.} explore the implications of the poorly-constrained rock-to-ice ratio within these planets to ask whether we should still be using the term `Ice Giants'; Mousis \textit{et al.} review recent ground-based measurements of atmospheric composition to constrain planetary origins; and Friedson discusses the possibilities for probing the Ice Giant interiors via seismology, following helioseismology techniques used in solar physics.  The Ice Giants may host exotic states of matter not found elsewhere, such as partially dissociated deep-water oceans \cite{11redmer} and superionic water ice mantles \cite{18millot}. New measurements of the internal structure and rotation (via detailed mapping of gravity and magnetic fields), in addition to determinations of the planetary temperatures and composition below the visible atmospheric layers, will shed new light on the origins of these two worlds, and on formation and evolution processes at work throughout our universe.

\subsection{Magnetospheres}

The magnetospheres of the Ice Giants are without parallel in our Solar System, with the substantial misalignments between the magnetic dipoles and planetary rotation axes creating highly unusual and time-variable interactions with the external solar wind.  The system is more complex at Uranus with the planet's high obliquity, creating a unique configuration with a strong helical structure near solstice, allowing us to test our understanding of planetary magnetospheric dynamics and evolution to the extremes \cite{20arridge}.  The uniqueness of these laboratories for astrophysical plasma processes was highlighted during the first day of the Discussion Meeting through several contributions, summarised in this special issue:  Soderlund \textit{et al.} explore the internal dynamo processes responsible for the magnetic field, and how they differ from other magnetised planets; Paty \textit{et al.} describe the processes shaping the Ice Giant Magnetospheres; and Lamy \textit{et al.} reveal how plasma and solar wind processes might be generating auroral phenomena.  With the exception of auroras, which can just about be imaged in the UV and near-IR from Earth, our understanding of the magnetospheres comes entirely from the \textit{in situ} measurements from the Voyager spacecraft, combined with subsequent magnetospheric modelling.  Magnetic field, plasma, and wave measurements from a future orbital mission will be essential to improve our understanding of these unique systems.

\subsection{Satellites and Rings}

The second day of the Discussion Meeting moved from the Ice Giant planets themselves, out into the wider systems represented by their satellites and rings.  Although the rings can be observed from ground- and space-based facilities (albeit at limited phase angles accessing only backscattered light and thermal emission) \cite{06depater_rings}, the lack of spatial resolution means that the satellites have been only points of light\footnote{Limited spatial resolution can be attained by measuring both the leading and trailing hemispheres of the unresolved satellites, and spectroscopy of the satellites have revealed insights into their surface composition \cite{18cartwright}.} since the Voyager observations.  Nevertheless, the precious and limited maps of their surface geology and chemistry, in addition to geophysical information about their bulk properties, have been explored in great detail.  The large satellites of Uranus (Miranda, Ariel, Umbriel, Titania and Oberon, Fig. \ref{system}) are considered to be a natural, primordial Ice Giant satellite system, with geological diversity revealing signs of past resurfacing, tectonics, and potential cryovolcanism.  Only the southern hemispheres of these moons have ever been seen, due to the southern-summer season in 1986.  A major outstanding question is whether tidal interactions could lead to the presence of liquid water oceans beneath their icy crusts \cite{06hussmann}, therefore extending the zone of potential habitability out to solar distances that were previously unimaginable.  In this special issue, Schenk and Moore explore the diverse geology of these satellites.


In contrast to the Uranian system, the Neptunian system appears to have been substantially altered by the presence of an interloper from the Kuiper Belt:  the large moon Triton, with its own atmosphere, active geology and plumes of nitrogen gas and dust, providing an intriguing connection between a future Neptune mission and the recent New Horizons exploration of Pluto and 486958 Arrokoth.  Triton is considered as a key ``Ocean World'' for future exploration \cite{19hendrix}, and may also harbour a subsurface ocean \cite{06hussmann}.  Beyond the major moons, almost nothing is known about the minor satellites of Uranus and Neptune, as they were too small and distant for even Voyager's cameras. Showalter \textit{et al.} explore the gravitational and source/sink connections between these small moons and the narrow and dusty rings (as a counterpoint to the extensive rings of Saturn), which were recently reviewed by Nicholson \textit{et al.} for Uranus \cite{18nicholson_rings} and de Pater \textit{et al.} for Neptune \cite{18depater_rings}.  The origin, evolution, and gravitational relationships of the Ice Giant rings may offer new insights into the forces shaping ring systems around planetary bodies.

\subsection{Astrophysical and Heliophysical Connections}

A future mission to the Ice Giants, as representatives of one of the most common outcomes of the planet formation process \cite{13fressin, 18fulton}, could provide a foundation stone for understanding the properties of distant, unresolved extrasolar Neptunes and sub-Neptunes\footnote{We note that the exoplanetary community could easily have substituted `Uranus` for `Neptune` in this planetary taxonomy with no change of meaning.}, as explained by Wakeford and Dalba in this special issue.  Any future robotic spacecraft visiting Uranus or Neptune would be operating during an explosion in new exoplanet discoveries (e.g., from ongoing exploitation of data from missions like JWST, CHEOPS, TESS, WFIRST and ARIEL), as well as new capabilities for atmospheric characterisation probing cooler temperature regimes.  The smaller radii of the Ice Giants compared to Jupiter means that they provide access to a different dynamical regime, accessing atmospheric phenomena (circulation, banding, vortices) on intermediate-sized worlds that are not available elsewhere \cite{19fletcher_V2050}.  

Furthermore, the complex orbital and magnetic characteristics of Uranus may be commonplace beyond our Solar System, making Uranus an ideal laboratory for exploring magnetospheric dynamos, dynamics, and transport within such a complex system \cite{19rymer}.   Closer to home, the magnetospheric configurations of Uranus/Neptune may also be relevant for understanding the Earth's palaeomagnetosphere during periods of geomagnetic reversal, potentially providing information on the solar wind driving of our own atmosphere/climate over geological time.  Finally, a mission cruising through the poorly-explored realm beyond Saturn would be sampling the heliosphere at great distances from the Sun, a new outpost for understanding how the solar wind influences planetary environments.   Cohen \& Rymer (this issue) describe some of the interdisciplinary opportunities offered by an ambitious Ice Giant mission, demonstrating how it would reach far beyond the planetary science community.

\section{Future Missions to the Ice Giants}

A return to the Ice Giants with an ambitious robotic spacecraft has long been recognised as the next natural step in our exploration of the Outer Solar System.  Following the legacy of the Cassini-Huygens exploration of the Saturn system, it is hoped that such a mission would combine orbital exploration of the planet, rings, satellites, and extensive magnetosphere, alongside \textit{in situ} atmospheric entry probe(s) and potentially landed elements on the icy moons (Fig \ref{system}).  A comprehensive and diverse instrument suite enables the tremendous interdisciplinary science opportunities described in the previous sections.  But the scope of the ambition is matched by the scale of the price tag, meaning that the excellence of the science is a \textit{necessary but not sufficient condition} for developing a large-class international ``flagship'' that could last for multiple decades (as explained in this special issue by Hammel).  

Mission concept developments are driven by (i) funding opportunities created by space agencies (e.g., the US decadal surveys\footnote{\url{https://science.nasa.gov/about-us/science-strategy/decadal-surveys}} in planetary science, and the European Cosmic Vision\footnote{\url{https://sci.esa.int/web/cosmic-vision}} and Voyage 2050\footnote{\url{https://www.cosmos.esa.int/web/voyage-2050}} programmes); and (ii) launch opportunities offered by gravity-assist trajectories (e.g., typically using Jupiter, with its 13-14 year synodic period with Uranus and Neptune) or new heavy-lift launchers.  If Jupiter Gravity Assist (JGA) is a requirement for the delivery of sufficient spacecraft mass into the system, then a recent NASA-ESA joint study described by Hofstadter \textit{et al.} \cite{19hofstadter} highlighted optimal launches to Neptune in 2029-30, and a wider window for Uranus in the early 2030s (although non-JGA trajectories are available \cite{14arridge}).  These would have missions arriving in the 2040s, as Uranus approaches northern autumnal equinox (2050) and Neptune reaches northern spring equinox (2046) \cite{19fletcher_V2050}.  Such a trajectory would also allow a spacecraft to visit other Solar System objects (such as Centaurs) during the cruise phase, as well as observing the solar wind conditions over a large range of heliocentric distances \cite{14arridge, 14masters, 19fletcher_V2050}.  Spilker \textit{et al.} (this issue) discuss some of the technological challenges associated with meeting this launch opportunity, including the challenge of powering the spacecraft via radioisotope thermoelectric generators, as recently reviewed by Ambrosi \textit{et al.} \cite{20ambrosi}.  

The Royal Society Discussion Meeting included contributions from the lead proposers of Ice Giant mission concepts over the past several decades.  Missions that could have taken advantage of JGA in 2019-20 were proposed in the early 21st century, such as the Argo mission that would have flown past Neptune and Triton in 2029, and then on to a Kuiper Belt Object \cite{09hansen_ARGO}.  More recent mission concepts have targeted the 2029-35 timeframe.  The previous US planetary decadal survey 2013-2022\footnote{\url{https://www.nap.edu/catalog/13117/vision-and-voyages-for-planetary-science-in-the-decade-2013-2022}} listed a Uranus Orbiter and Probe\footnote{Both Uranus and Neptune were considered during the decadal survey deliberations, Hubbard \textit{et al.}: \url{https://solarsystem.nasa.gov/studies/225/ice-giants-decadal-study/} and Marley \textit{et al.}: \url{https://solarsystem.nasa.gov/studies/226/jpl-rapid-mission-architecture-neptune-triton-kbo-study-final-report/}} as its third highest priority, after the newly-named Mars Perseverance (2020) rover mission and Europa Clipper mission.  For Europe's Cosmic Vision, Arridge \textit{et al.} proposed medium-class (500 MEur) Uranus-orbiting spacecraft in 2010 and 2014, based on heritage from Mars Express and Rosetta \cite{12arridge}; whilst Mousis \textit{et al.} proposed a European atmospheric entry probe design for Saturn \cite{16mousis_hera} and the Ice Giants \cite{18mousis}.  A 2013 call for large-class mission themes within ESA's Cosmic Vision resulted in proposals for a Uranus orbiter with probe \cite{14arridge}, a mission for Neptune and Triton \cite{14masters}, and a concept for dual orbiters of both worlds \cite{14turrini}.  The resulting strong support of ESA's Senior Survey Committee \cite{13ssc} led to European participation in a NASA-led science definition team (2016-17) that explored different Ice Giant mission architectures more closely \cite{19hofstadter}, and allowed ESA to consider (2018-19) a palette of potential contributions to a US-led mission, including secondary orbiters and atmospheric probes, as reviewed by Simon \textit{et al.} \cite{20simon_icegiant}.  It may also be possible to develop missions that visit both Uranus and Neptune as part of an integrated strategy in the forthcoming decade \cite{18simon_icegiant} - for example, a Uranus flyby that could continue on to explore the distant Kuiper Belt, in addition to a flagship-class Neptune orbiter with atmospheric probe.  Finally, novel Ice Giant mission concepts have continued to emerge from teams taking part in planetary science summer schools at NASA/JPL (e.g., MUSE \cite{14saikia_MUSE}, OCEANUS \cite{18elder_OCEANUS}, and QUEST \cite{20jarmak_QUEST}) and the Alpach Summer School (MUSE \cite{15bocanegra}).

To date, none of these concepts have progressed to become a formal new mission, but they have been essential in preparing the ground for the coming decade.  At the time of writing, community input and mission concepts have been sought for both ESA's Voyage 2050 programme \cite{19fletcher_V2050} (covering the period 2025-2050, following on from ESA's Cosmic Vision 2025) and the next US planetary decadal survey (2023-32).  In addition, US scientists are studying a Triton flyby mission (Trident, \cite{19prockter_trident}) as a potential Discovery-class mission, and a larger flagship-class Neptune orbiter and probe (Odyssey\footnote{\url{https://www.hou.usra.edu/meetings/exoplanets2020/presentations/Rymer.pdf}}).  The 2020 Royal Society Discussion Meeting emphasised the high desirability of international partnership on these missions, but that this requires closer alignment of national and international mission opportunities (i.e., inter-agency collaboration).  Whether that can happen in time for the 2029-35 JGA opportunities remains to be seen.

\section{Conclusion: The Next Decade}

This special issue, alongside the plethora of scientific reviews and mission concepts outlined in this article, demonstrate the enthusiasm and momentum for an ambitious, paradigm-shifting international mission to the Ice Giants in the coming decades.  This will be the first dedicated mission to the last-remaining class of major planets to be explored, completing humankind's first reconnaissance of the eight planets.  The potential for discovery is vast, and we urge our national and international space agencies to take up this challenge.  We hope that the 2020 Royal Society Discussion Meeting will serve as a memorable stepping stone towards meeting this ambitious goal for the next generation of planetary explorers.  

\vskip6pt







\enlargethispage{20pt}


\dataccess{No new data was used or generated in the preparation of this review article.}

\aucontribute{LNF was the lead proposer for the Royal Society Discussion Meeting on the \textit{Future Exploration of Ice Giant Systems}, and wrote this article.  All other authors were co-proposers for the Discussion Meeting, and provided additional content and editorial corrections for this manuscript.  All authors read and approved the manuscript.}

\competing{The author(s) declare that they have no competing interests.}

\funding{Fletcher, Arridge and Masters were supported by Royal Society Research Fellowships.  Fletcher acknowledges support from a European Research Council Consolidator Grant (under the European Union's Horizon 2020 research and innovation programme, grant agreement No 723890) at the University of Leicester.}

\ack{The authors wish to thank the Royal Society for their tremendous assistance in hosting this Discussion Meeting in January 2020.}

\disclaimer{N/A.}

\bibliographystyle{RS.bst} 
\bibliography{references.bib}







\end{document}